\documentclass[aps,reprint,prl,superscriptaddress]{revtex4-1}
\usepackage{graphics,amsmath,epstopdf}
\draft 
\usepackage[bookmarks=false]{hyperref}
\begin{document}

\title{Electrically driving nuclear spin qubits with microwave photonic bangap resonators} 

\author{A. J. Sigillito}
\email[]{asigilli@princeton.edu}
\affiliation{Department of Electrical Engineering, Princeton University, Princeton, New Jersey 08544, USA}

\author{A. M. Tyryshkin}
\affiliation{Department of Electrical Engineering, Princeton University, Princeton, New Jersey 08544, USA}

\author{T. Schenkel}
\affiliation{Accelerator and Fusion Research Division, Lawrence Berkeley National Laboratory, Berkeley, California 94720, USA}

\author{A. A. Houck}
\affiliation{Department of Electrical Engineering, Princeton University, Princeton, New Jersey 08544, USA}

\author{S. A. Lyon}
\affiliation{Department of Electrical Engineering, Princeton University, Princeton, New Jersey 08544, USA}


\date{\today}


\begin{abstract}

 The electronic and nuclear spin degrees of freedom for donor impurities in semiconductors form ultra coherent two-level systems that are useful for quantum information applications. Spins naturally have magnetic dipoles, so alternating current (AC) magnetic fields are frequently used to drive spin transitions and perform quantum gates. These fields can be difficult to spatially confine to single donor qubits so alternative methods of control such as  AC electric field driven spin resonance are desirable. However, donor spin qubits do not have electric dipole moments so that they can not normally be driven by electric fields. In this work we challenge that notion by demonstrating a new, all-electric-field method for controlling neutral $^{31}$P and $^{75}$As donor nuclear spins in silicon through modulation of their donor-bound electrons. This method has major advantages over magnetic field control since electric fields are easy to confine at the nanoscale. This leads to lower power requirements, higher qubit densities, and faster gate times. We also show that this form of control allows for driving nuclear spin qubits at either their resonance frequency or the first subharmonic of that frequency, thus reducing device bandwidth requirements. Interestingly, as we relax the bandwidth requirements, we demonstrate that the computational Hilbert space is expanded to include double quantum transitions, making it feasible to use all four nuclear spin states to implement nuclear-spin-based qudits in Si:As. Based on these results, one can envision novel high-density, low-power quantum computing architectures using nuclear spins in silicon.
\end{abstract}

\maketitle 

\section{Introduction}

Ever since Feynman's seminal paper envisioning quantum computers \cite{Feynman1982}, physicists have dreamt of the ability to simulate complex quantum systems. In the 90's, when quantum algorithms were discovered that could outperform their best known classical alternatives \cite{Grover1996, Shor1999}, the interest in quantum computing redoubled. Soon after, it was shown that quantum computers could be realized using the electronic and nuclear spins  of donors in silicon \cite{Kane1998} which have long coherence times \cite{Tyryshkin2012,Saeedi2013,Morton2008, Petersen2016} and are intrinsically compatible with industrial semiconductor processing. Because of their smaller gyromagnetic ratios, nuclear spins are more difficult to manipulate than electron spins and are often considered too slow for quantum information processing, but could be suitable as a quantum memory \cite{Morton2008}. In this work we demonstrate a new, scalable method for controlling nuclear spins that should allow one to perform rapid manipulations of nuclear spins in silicon. Using coplanar photonic bandgap resonators, we drive Rabi oscillations on nuclear spins using exclusively \textit{electric} fields by employing the donor-bound electron as a quantum transducer, much in the spirit of recent work with single-molecule magnets \cite{Thiele2014}. Electric control has major advantages over magnetic control since electric fields are easy to spatially confine at nanometer length scales. The field confinement leads to lower power requirements, higher qubit densities, and faster gate times. We also show here that electric field control allows for driving spin qubits at either their resonant frequency or the first subharmonic of that frequency, thus reducing device bandwidth requirements. Finally, we show that double quantum transitions can be driven which opens up a richer computational manifold and makes it feasible to implement nuclear spin based qudits using $^{75}$As donors \cite{Bartlett2002}.

In recent years, schemes for all-electrical control of donor spin qubits have been proposed \cite{De2009, Tosi2015} but no experimental demonstrations have been reported. Some success has been shown in fundamentally different spin systems including defects in SiC \cite{Klimov2014}, quantum dots \cite{Pioro2008, Laird2009, Petersson2012}, and single molecule magnets \cite{Thiele2014}, but this work represents the first demonstration of electrically driven nuclear magnetic resonance(EDNMR) for donor spins in silicon. This material system is particularly attractive since it already boasts record coherence times \cite{Morton2008, Saeedi2013, Petersen2016}, and atomically precise lithography techniques that can truly benefit from electrical control are becoming mature\cite{Fuechsle2012, Scappucci2011}.

We find that there are two distinct mechanisms which lead to EDNMR depending on the donor species. $^{31}$P donor nuclei are driven through modulation of the electronic orbital states through the spin-orbit Stark shift \cite{Bradbury2006,SigillitoStark2015,SigillitoGeStark}. This tilts the direction of the quantization axis of the electronic spins and induces effective anisotropy in the hyperfine interaction. $^{75}$As is subject to the same form of control but we find that the $^{75}$As Rabi frequencies are too large for this effect to be responsible. Electric field modulation of the quadrupolar coupling is likely responsible for the EDNMR in $^{75}$As.

\section{Experimental Methods}

In this work, we make use of the hyperfine interaction to read out the nuclear spin state ($m_{I}$) using the Davies electron-nuclear double resonance (ENDOR) technique \cite{Davies1974, Tyryshkin2006}. In this measurement, one probes the electron spin resonance (ESR) transitions while simultaneously performing nuclear magnetic resonance. The ESR transition intensity depends on the nuclear spin state, so by performing conventional ESR on the donor electron spin, one also obtains $m_{I}$. This technique therefore requires both microwave magnetic ($\vec{B}_{1}$) and radio frequency magnetic ($\vec{B}_{2}$) fields. In this study, to electrically probe nuclei, we also require RF electric fields ($\vec{E}_{2}$) fields. To maintain a suitable signal-to-noise ratio (SNR), a high quality factor microwave resonator is used. 

\begin{figure}[h]
	\includegraphics{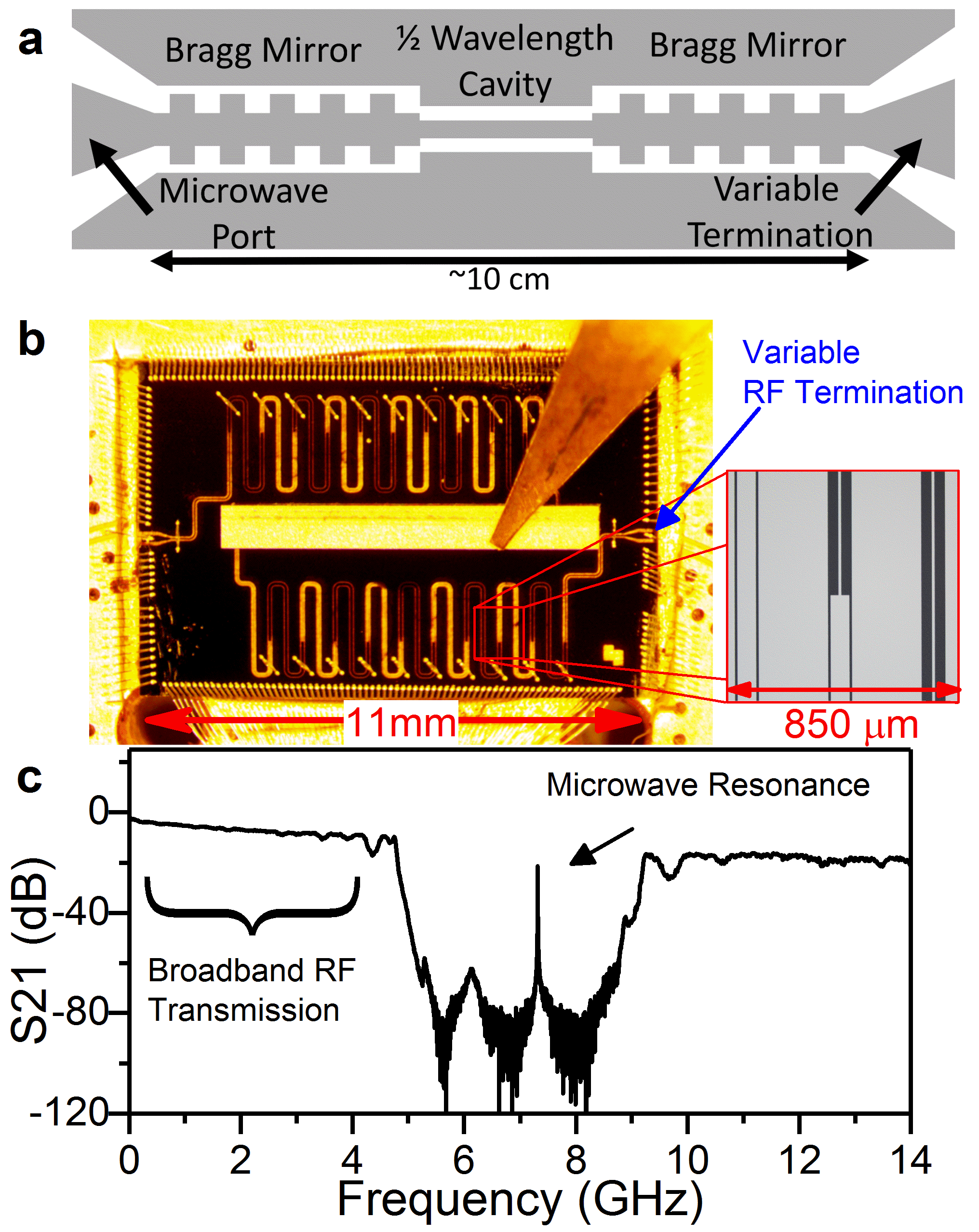}
	\caption{\label{fig:fig1} (a) Cartoon schematic of a photonic bandgap resonator. The left port of the device is used for microwave excitation and readout and the right port can be terminated to select whether RF electric or magnetic fields are present in the device. An optical micrograph of an actual device is shown in (b) with a silicon sample (bright rectangular feature) mounted using a phosphor bronze clip. The serpentine structures above and below the sample are the Bragg mirrors. The inset shows a zoomed in view of an impedance step. The microwave transmission through this structure is shown in (c) for the device in a magnetic field of 250 mT at a temperature of 1.9 K. The photonic bandgap spans the frequency range between 4.5 GHz and 9 GHz with nearly lossless transmission below 4 GHz. The resonance appears at 7.3 GHz with a Q factor of 20000. The loss outside of the bandgap is due to the coaxial test cables which were not calibrated out.} 
\end{figure}

Commercial ENDOR resonators exist, but they require large powers (making them incompatible with ultra-low temperature measurements) and are designed to provide RF magnetic, and not electric fields. This led us to develop superconducting coplanar photonic bandgap (PBG) resonators which allow broadband RF and microwave transmission above and below a lithographically defined photonic bandgap \cite{Yun1999, Liu2016, Yablonovitch1987, Bronn2015}. A schematic of the device is shown in Fig.~\ref{fig:fig1}(a). The bandgap is constructed by periodically alternating the impedance of a superconducting CPW transmission line to form a one-dimensional microwave Bragg grating as shown schematically in Fig.~\ref{fig:fig1}(a). By incorporating a 1/2 wavelength defect in the photonic bandgap, the device supports a resonant mode which can be used for ESR. Equivalently this structure can be thought of as two discrete Bragg mirrors defining the boundaries of a half wavelength cavity \cite{iga2008}. The sample is located above the cavity region of the device. This resonator design has a continuous center conductor which is isolated from the ground plane and allows for easy application of DC voltage or current biases. These devices require only one layer of lithography and will be convenient for other areas of quantum information processing and ESR. Resonator design considerations are outlined in the Supplementary Information. 

These resonators have a built-in feature which allows us to easily select whether electric or magnetic RF fields are present in the sample. The RF frequencies used in this work have a wavelength that is large compared to the scale of the device and are unperturbed by the photonic bandgap (since they lie well below the gap). We can therefore set up RF standing waves by terminating the transmission line at the output port of the device (labeled "variable termination" in Fig.~\ref{fig:fig1}(a)). A high impedance (open) termination is used to enhance $\vec{E}_{2}$ whereas a low impedance (shorted) termination enhances $\vec{B_{2}}$ in the sample. Due to the finite size of the device, one can never fully suppress the $\vec{E}_{2}$ and $\vec{B}_{2}$ fields but we estimate that the residual undesired field amplitudes are reduced by at least a factor of 50 in the sample. The microwave magnetic field, $\vec{B}_{1}$, has a wavelength that is set by the $\lambda/2$ section of the device and is well confined by the two Bragg mirrors. It is unperturbed by the termination off-chip so that we can select between $\vec{E}_{2}$ and $\vec{B}_{2}$ in the device without changing $\vec{B}_{1}$ or the ensemble of spins probed by the ESR.

The resonators used in this work were patterned in a 50 nm thick Nb film e-beam evaporated on the surface of a C-plane sapphire wafer. The structures were defined using optical lithography and SF$_{6}$ plasma etching as previously described \cite{Sigillito2014, Malissa2013}. These resonators can be patterned directly on the silicon sample to offer enhanced sensitivity, but to ensure that the spin signal only comes from the 1/2 wavelength defect region of the resonator (and not spins within the Bragg mirrors), the sample was clipped to the surface of the resonator using a phosphor bronze spring as shown in Fig.~\ref{fig:fig1}(b). This particular device has five periods of Bragg mirror on either side of the half wavelength defect. Each period consists of both a high impedance (95 $\Omega$, 4 mm long) and a low impedance (30 $\Omega$, 4 mm long) strip of waveguide. The cavity is 6 mm long with a 10 $\mu m$ wide center pin and gap. The RF termination is defined by either leaving the output port floating, or by shorting it to ground using aluminum wirebonds.

The device was cooled to 1.9 K in a pumped helium cryostat equipped with a rotatable sample holder. This allows for \textit{in-situ} alignment of the device with an externally applied magnetic field $\vec{B}_{0}$ \cite{Malissa2013}. With $B_{0} = $ 250 mT applied in the plane of the Nb, the microwave transmission spectrum was measured and is plotted in Fig.~\ref{fig:fig1}(c). The photonic bandgap gives about 80 dB of attenuation from 4.5 - 9 GHz and the microwave resonance appears at 7.3 GHz. The resonator is slighly undercoupled and has a temperature-limited quality factor of approximately 20000. The spin sensitivity of this resonator was determined to be $5 \times 10^{6}$ spins per shot at 2 K using phosphorus doped $^{28}$Si (800 ppm $^{29}$Si). This is on par with other planar resonators \cite{Sigillito2014, Malissa2013} and could be further improved by incorporating quantum-limited parametric amplifiers \cite{Bienfait2016sensitivity, Eichler2016}.

\begin{figure}[h]
\includegraphics{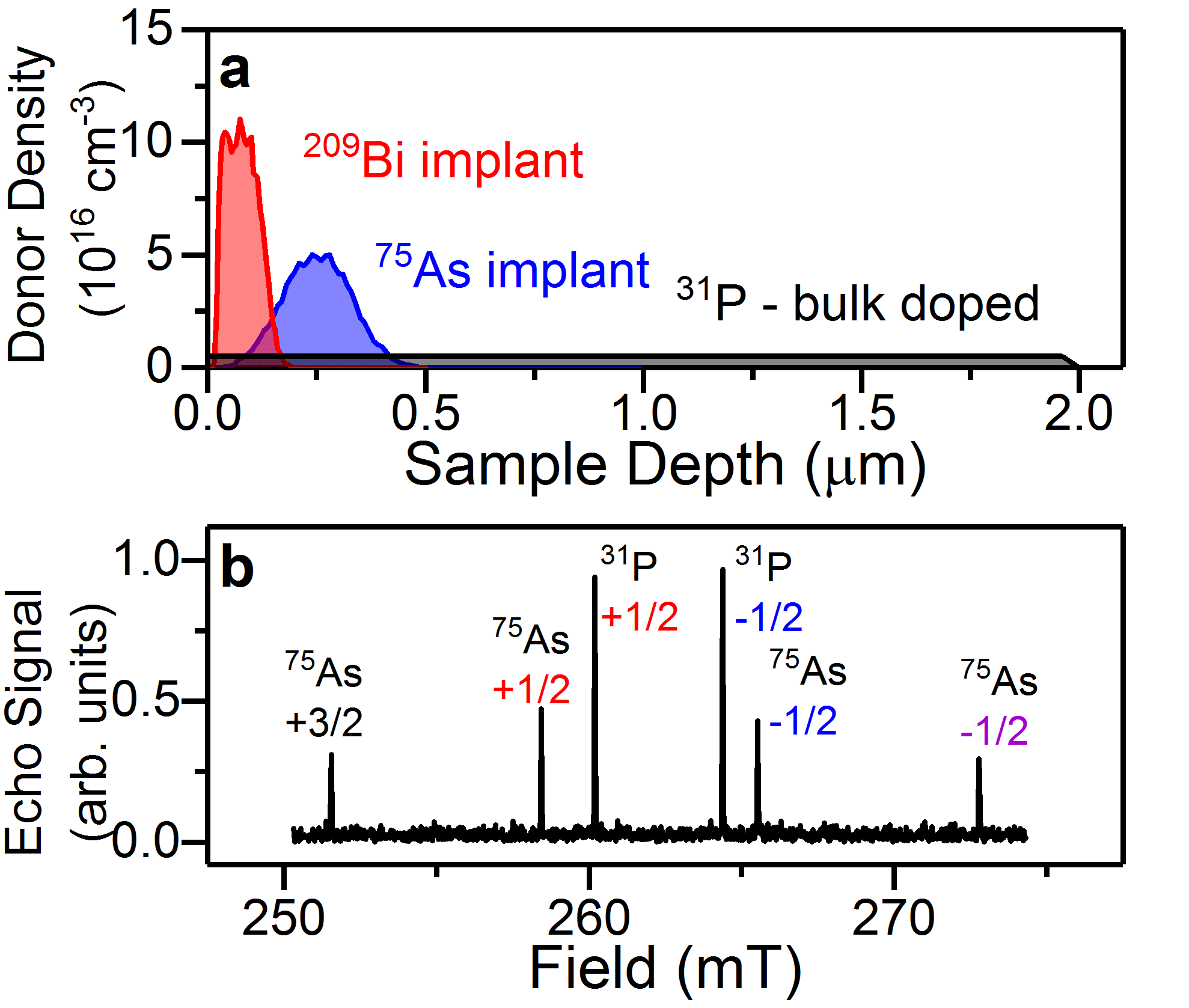}
\caption{\label{fig:fig2} The sample consists of a $^{31}$P doped $^{28}$Si epitaxial layer which has been implanted with $^{209}$Bi and $^{75}$As as shown in (a). An echo detected field sweep spectrum is shown in (b) resolving the two $^{31}$P and the four $^{75}$As hyperfine lines. The hyperfine lines are labeled by their nuclear spin projections with colors matching the data in Fig.~\ref{fig:fig3}. These data were taken at 1.9 K with a resonator frequency of 7.3 GHz.} 
\end{figure}

The sample used throughout this work consists of a 2 $\mu$m isotopically enriched $^{28}$Si epitaxial layer grown on a high resistivity p-type substrate. The epi-layer was grown to have 5$\times$10$^{15}$ $^{31}$P/cm$^{3}$ and the sample was ion implanted with $^{209}$Bi and $^{75}$As. After implantation, the donors were activated by annealing the sample in a N$_{2}$ atmosphere at 800$^{\circ}$C for 20 minutes \cite{Weis2012}. The simulated implantation profiles are shown in Fig.~\ref{fig:fig2}(a) \cite{SRIM}. Two-pulse Hahn echo measurements were performed at 1.9 K and an echo-detected field swept spectrum is shown in Fig.~\ref{fig:fig2}(b), revealing the $^{31}$P and $^{75}$As hyperfine lines. Using pulsed spin counting techniques, we estimate the $^{209}$Bi activation to be about 50\% whereas the $^{75}$As donors are fully activated. Because the $^{209}$Bi signal is very weak (due to low donor activation and a large nuclear spin, 9/2), the ENDOR experiments were only performed on the $^{31}$P and $^{75}$As donors.

In the presence of $\vec{B}_{1}$ inhomogeneity, one can measure entirely different subensembles of spins subject to different RF electric or magnetic fields by varying the microwave power\cite{Sigillito2014}. It was therefore important to calibrate $\vec{B}_{1}$ before every ENDOR experiment by performing two-pulse Hahn echo experiments as a function of microwave power. The electric and magnetic field distributions are well known\cite{Malissa2013,SigillitoStark2015} and are plotted in the supplementary information. It has been shown that inhomogeneity in $\vec{B}_{1}$ can be overcome by using adiabatic (BIR-WURST) pulses \cite{Sigillito2014}. In the supplementary information we demonstrate that they also overcome $\vec{E_{2}}$ inhomogeneity. These pulse shaping techniques make PBG resonators useful for complex ENDOR experiments requiring high fidelity manipulations. We measure Rabi frequencies in the following experiments, which were thus conducted using rectangular pulses.

Prior to every experiment, the spins were prepared in thermal equilibrium using a combination of RF and optical pulses as described in \cite{Tyryshkin2006} since nuclear spin relaxation times are long at these temperatures.

\section{Results}

Davies ENDOR experiments were first performed using only RF magnetic fields (shorted device termination). The ENDOR spectra for all four of the $^{75}$As donor hyperfine lines are plotted in Fig.~\ref{fig:fig3}(b) but the experiments were also performed on the $^{31}$P donors as shown in the supplementary information. Only the magnetic dipole allowed transitions could be resolved in this configuration. Those are $\Delta m_{s}=0, \Delta m_{I}=\pm 1$, with $m_{s}$ and $m_{I}$ being the electronic and nuclear spin projections, respectively. 

The device was reconfigured to have $\vec{E}_{2}$ fields in the sample (open termination) and the measurements were repeated with the results displayed in Fig.~\ref{fig:fig3}(c) ($^{31}$P data available in the supplementary information). In addition to the allowed ENDOR transitions, several additional transitions appeared in the $^{75}$As spectra and are denoted by the arrows. These very narrow transitions occur at exactly half the frequency of forbidden double quantum transitions ($\Delta m_{s}=0, \Delta m_{I} = 2$). The single quantum transitions are power broadened in this plot, since power was optimized for the double quantum transitions. Transitions were also observed at subharmonics of the allowed transition frequencies and are shown in Fig.~\ref{fig:fig3}(d).

The double quantum transitions do not exist for $^{31}$P donors since they have nuclear spin-1/2. EDNMR was observed at the fundamental and subharmonic transitions frequencies for $^{31}$P, but it was noticed that $^{31}$P donors require more RF power than $^{75}$As donors. To quantify the difference, two dimensional EDNMR measurements of the Rabi nutation were conducted on both donors. These experiments used the standard Davies ENDOR pulse sequence but varied the RF pulse length and power. The data for the subharmonic transitions are plotted in Fig.~\ref{fig:fig4} (a-b) and the Rabi data for the fundamental transitions are shown in the supplementary information. From the data, it is clear that the arsenic donors respond over an order of magnitude more strongly to the electric fields (shorter RF pulses are necessary); indicating that different mechanisms may be responsible for the EDNMR in these two donors.

\begin{figure}[h]
	\includegraphics{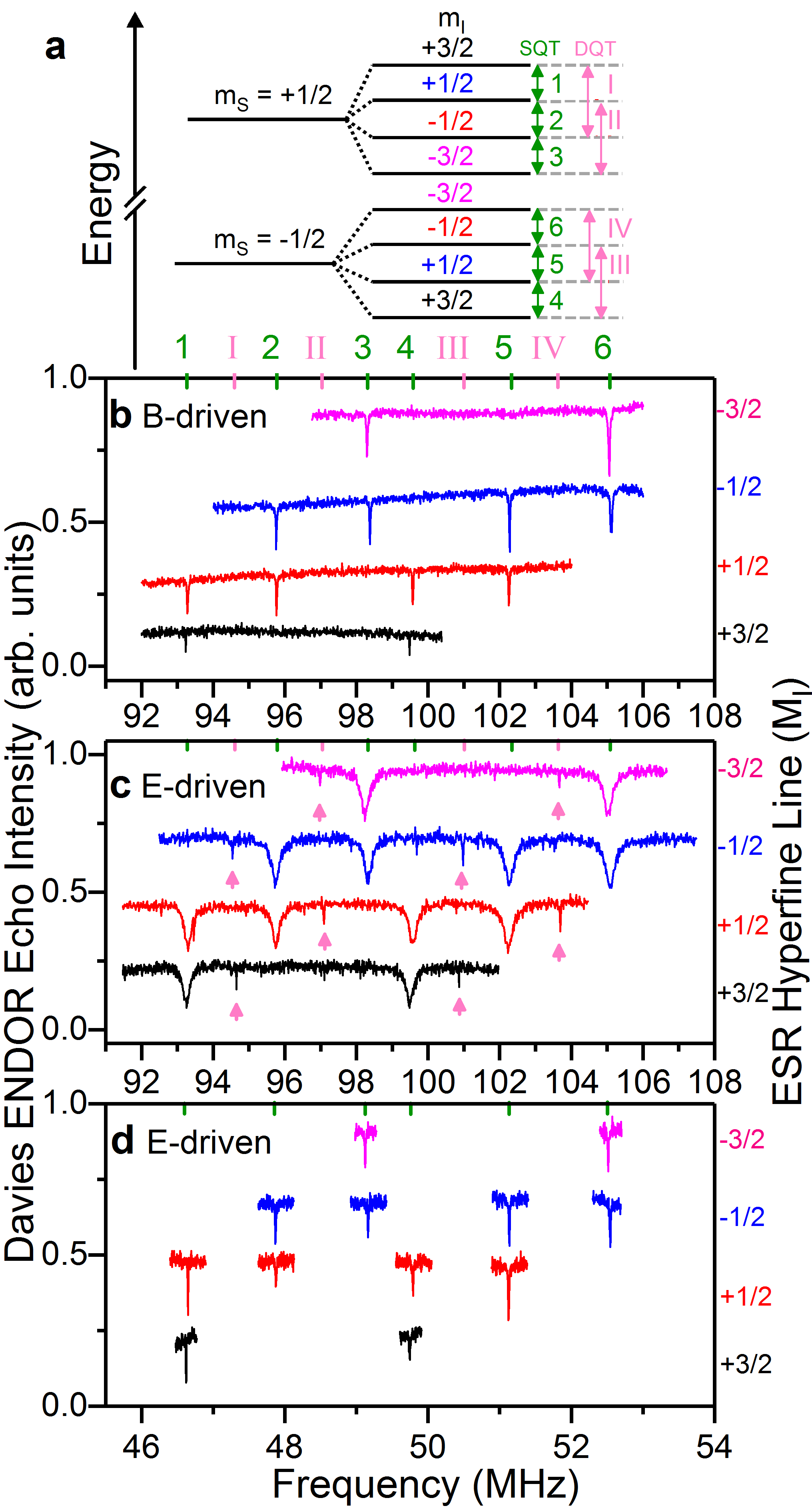}
	\caption{\label{fig:fig3}(a) Energy level diagram illustrating the electronic Zeeman ($m_{S}$) and nuclear hyperfine ($m_{I}$) splittings for $^{75}$As donors in Si. The single (SQT) and double (DQT) quantum transitions are labeled using the green numbers and pink numerals, respectively. These labels are also used in (b-d). The Davies ENDOR spectra measured using magnetic (b) and electrical (c-d) RF pulses are plotted. The magnetically driven ENDOR spectra shows the six SQTs whereas the electrically driven spectra (c) reveals both the SQTs and DQTs. Electrically driven ENDOR also resolves transitions at subharmonics of the SQTs as shown in (d). The SQTs in (b) are power broadened.} 
\end{figure}

To verify that residual RF magnetic fields (due to the finite length of the device) can not be responsible for the EDNMR, $\vec{B}_{2}$ was calibrated in the device using a Rabi-nutation experiment and the results were compared against the EDNMR data for both $^{31}$P and $^{75}$As. We found one would need 300 times more power for the residual $\vec{B}_{2}$ fields to account for the EDNMR.

To ensure that the subharmonic transitions were not artifacts driven by second harmonics generated in the RF source, the output of the RF source was fed directly into a spectrum analyzer. We observed that in the worst case configuration, a second harmonic was present and attenuated by 35 dB compared to the fundamental harmonic. To further suppress this second harmonic, a set of seventh order Butterworth low pass filters (Crystek CLPFL) were used in every experiment, adding 35-50 dB of attenuation. Given the more than 70 dB power difference between the first and second harmonics, we are confident that the observed subharmonic ENDOR transitions are not due to harmonics generated from the RF source.

\section{Discussion}

This is the first demonstration of electrically driven NMR for donors in silicon. We have identified two mechanisms that likely lead to the observed EDNMR, but more theoretical and experimental work will be needed to confirm our explanation.

To understand the EDNMR, we turn to the spin Hamiltonian common to group V donors in silicon. This is given by

\begin{equation} 
H/h = \beta \vec{B}_{0} \cdot \hat{g}_{e} \cdot \vec{S} + \vec{S}\cdot \hat{A} \cdot \vec{I} - \beta_{n} g_{n} \vec{B}_{0} \cdot \vec{I} + \vec{I} \cdot \hat{Q} \cdot \vec{I}
\end{equation}

where $\beta$ is the Bohr magneton, $\hat{g}_{e}$ is the electron gyromagnetic tensor, $\vec{S}$ is the electronic spin, $\hat{A}$ is the hyperfine tensor, $\vec{I}$ is the nuclear spin, $g_{n}$ is the nuclear g-factor, $\beta_{n}$ is the nuclear magneton, and $\hat{Q}$ is the nuclear quadrupole coupling tensor. The terms in the spin Hamiltonian that are sensitive to electric fields are the electronic Zeeman ($\hat{g}_{e}$), hyperfine ($\hat{A}$), and quadrupolar ($\hat{Q}$) tensors. Because EDNMR is observed for both $^{75}$As and $^{31}$P (which has no quadrupole moment), we first neglect quadrupolar effects.

Both $\hat{A}$ and $\hat{g}_{e}$ can be modulated through the hyperfine and spin-orbit Stark effects, respectively. These effects are quadratic to first order due to inversion symmetry at the donor site, but linear terms can arise from strain\cite{Feher1961,Pica2014}. We therefore expect to drive transitions at both the electric field frequency, $f$, and $f/2$ (since sin$^{2}(f) \propto$ cos($2f$)). Similar subharmonic transitions have been observed for electrically driven spin resonance in quantum dots\cite{Laird2009,Romhanyi2015}. Since the fundamental transition (at $f$) is strain dependent, we will restrict our discussion to the subharmonic transition which should be more robust against sample-specific strains.

Spin transitions cannot be driven solely by modulation of an isotropic hyperfine interaction due to the disparity in the electronic and nuclear precession frequencies. Any transition matrix elements involving $A_{XX}$ and  $A_{YY}$ terms average out in the rotating wave approximation and $A_{ZZ}$ terms can not drive spin rotations. We therefore require an anisotropic hyperfine interaction with $A_{ZX}$ terms to drive nuclear spins. To find the source of the anisotropy, we turn to the spin-orbit Stark shift.

We can compute the electric field modulation of $\hat{g}$ using the multivalley effective mass theory of \cite{Feher1961} and the experimental Stark shift values from \cite{SigillitoStark2015} and \cite{Pica2014} as outlined in the supplementary information. We find that the spin-orbit Stark shift directly modulates the quantization axis of the electron spin such that electron and nuclear spins are quantized along different axes. RF electric fields then lead to RF modulation in the hyperfine field, as seen by the nuclear spin, which can lead to nuclear spin rotations.

\begin{figure}
\includegraphics{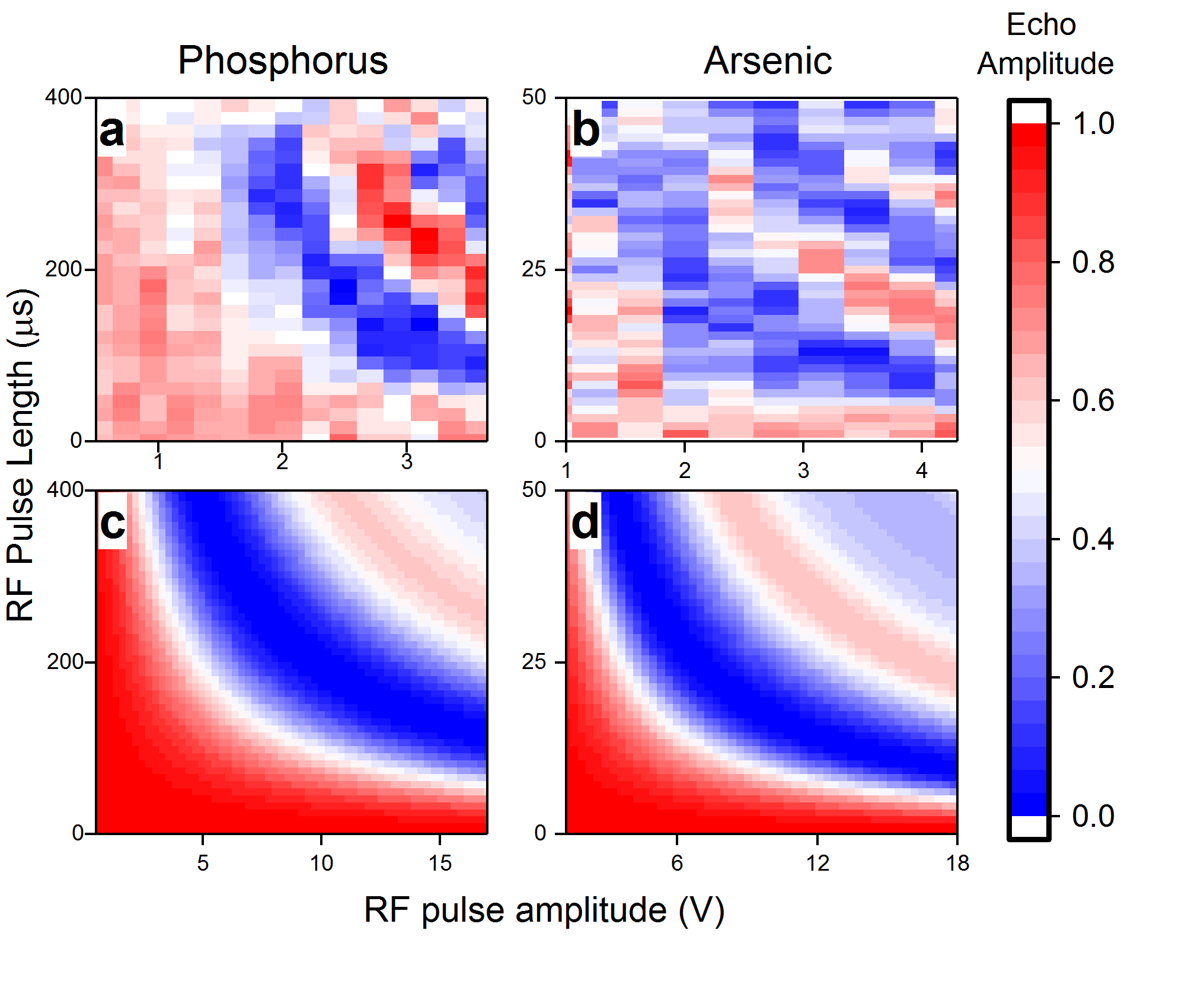}
\caption{\label{fig:fig4} Rabi oscillations are recorded as a function of RF voltage amplitude for $^{31}$P (a) and $^{75}$As (b) subharmonic transitions. The $^{75}$As transition is at 46.5 MHz and the $^{31}$P transition is at 54 MHz. The simulated plots (c-d) show similar dependences to the data, but larger RF amplitudes must be assumed indicating that our models underestimate the transition frequencies by a factor of 3-4. The phosphorus simulation takes into account g-tensor modulation leading to an anisotropic hyperfine coupling whereas the arsenic simulation also takes into account quadrupolar modulation. All data were taken at 1.9 K in a magnetic field of 250 mT.} 
\end{figure}

To test this mechanism against the experiment, we developed a model that simulates the Rabi-nutation experiments in our device. This model accounts for rotation angle errors in both the ESR and NMR pulses, spin-resonator coupling, inhomogeneity in the RF and microwave fields, and the implant profile of the donors. This simulation takes into account valley repopulation for electric fields in the (100) crystallographic directions (the dominant effect) but neglects the ''single valley" effect described in \cite{Feher1961}. We find reasonable agreement with the experimental data for the phosphorus donors as shown in Fig.~\ref{fig:fig4}(c). Note that the simulation is plotted on a different voltage scale indicating a 4$\times$ discrepancy between the simulation and data, which is reasonable given our necessarily rough estimates of the parameters. However, the equivalent $^{75}$As simulation (shown in supplementary information) would require a 40$\times$ larger voltage which implies that another mechanism must dominate.

The only significantly different term in the $^{75}$As and $^{31}$P Hamiltonians is the quadrupolar coupling, so this is a possible source for the discrepancy in the EDNMR Rabi frequencies. It is difficult to determine transition frequencies for quadrupolar modulation, due to uncertainties in screening potentials from inner shell electrons and no studies have reported electric field induced modulation of the quadrupolar interaction. There have, however, been several recent reports of quadrupolar shifts for $^{75}$As\cite{Franke2015,Franke2016neutral} and $^{209}$Bi \cite{Pla2016} donors in silicon subject to strain. By comparing the strain-induced hyperfine splitting measured in \cite{Franke2016neutral} to the hyperfine Stark data of \cite{Pica2014}, we can approximately scale the experimentally measured quadrupolar shifts in \cite{Franke2016neutral} to correspond to the electric fields we apply in our experiments. By repeating the Rabi-nutation experiment simulations while including the modulation of the quadrupolar interaction, the Rabi frequencies are enhanced by more than a factor of 10 as shown in Fig.~\ref{fig:fig4}(d). We find reasonable agreement to our data, again given the uncertainties in the parameters. We therefore conclude that the quadrupolar interaction is most likely responsible for the spin transitions in $^{75}$As, however more theoretical work is warranted.

For the largest RF electric fields applied in this work, the average Rabi frequency is approximately 70 kHz for the fundamental transition and 60 kHz for its subharmonic. These applied fields are a factor of 10 below the donor ionization threshold \cite{Lo2014}, indicating that MHz frequency EDNMR manipulations should be possible in unstrained Si. The fundamental transition Rabi frequencies depend on strain, suggesting that strain engineering should allow one to achieve even higher Rabi frequencies and the ultimate limit is unknown.

\section{Conclusion}

In summary, we have demonstrated the use of coplanar photonic bandgap microresonators to perform low temperature, high sensitivity ESR and ENDOR on $^{28}$Si:P and $^{28}$Si:As. We demonstrate for the first time all-electric-field control of donor nuclear spins in silicon for $^{75}$As and $^{31}$P donors using these structures. The EDNMR appears to arise from two distinct physical mechanisms. First, electric fields modulate the $g$-tensor to induce an effective anisotropic hyperfine interaction which appears to drive the $^{31}$P nuclear spin transitions. Secondly, for $^{75}$As, the quadrupolar interaction can be modulated to rotate nuclear spins. These experiments probe new physical mechanisms that manipulate nuclear spins which appear to depend on not only the electronic orbital structure, but also the interaction of the inner shell electrons with the donor-bound electron. As such, this should lead to new physical insights in the donor electron system.

Our technique for controlling nuclear spins has several advantages over magnetic control. It relaxes power requirements since voltages rather than currents are used, and allows for high density, individually addressable arrays of donor nuclear spins since electric fields are more easily confined than magnetic fields. Since we are able to drive spins at subharmonics of their resonance frequencies and at subharmonics of double quantum $\Delta m_{I} = 2$ transitions, our electric field control method substantially reduces the bandwidth requirements for quantum devices while simultaneously expanding the computational Hilbert space. From these results, one can envision new quantum computing architectures based on donor nuclear spins in silicon. These techniques should extend to other material systems with long coherence times such as donors in germanium\cite{SigillitoGe2015} which offer a four-order-of-magnitude enhancement in the spin-orbit Stark shift \cite{SigillitoGeStark,Pica2016}. The larger Stark effect should translate into significantly faster EDNMR gates.



%
%

%

\begin{acknowledgments}
We thank G. Pica, J.~J.~L. Morton, and P. Bertet for stimulating discussions.
Work at Princeton was supported by the NSF through the Princeton MRSEC (Grant No. DMR-01420541) and by the ARO (Grant No. W911NF-13-1-0179). Work at LBNL was performed under the auspices of the U.S. Department of Energy under Contract No. DE-AC02-205CH11231.
\end{acknowledgments}

\providecommand{\noopsort}[1]{}\providecommand{\singleletter}[1]{#1}%

\end{document}